[White Paper]

# W3ID: A Quantum Computing-Secure Digital Identity System Redefining Standards for Web3 and Digital Twins

Joseph Yun[1,2,*], Eli Lifton[1], Eunseo Lee[1], Yohan Yun[1], Abigail Song[1], Joshua Lee[1], Cristian Jimenez-Bert[1], Benedict Song[1], Yejun Lee[1], Alex Seo[1] and Sijung Yun[1,2,*]

1: Predictiv Care, Inc., Mountain View, CA, USA

2: Johns Hopkins University, Baltimore, MD, USA

*Correspondences: Joseph Yun, jyun21@jh.edu, Sijung Yun, syun16@jhu.edu

All authors worked for Predictiv Care, Inc. for this project

Abbreviations: Web3 standard meeting universal digital ID (W3ID), Digital Object Identifiers (DOIs)

## Abstract

The rapid advancements in quantum computing present significant threats to existing encryption standards and internet security. Simultaneously, the advent of Web 3.0 marks a transformative era in internet history, emphasizing enhanced data security, decentralization, and user ownership. This white paper introduces the **W3ID**, an abbreviation of Web3 standard meeting universal digital ID, which is a Universal Digital Identity (UDI) model designed to meet Web3 standards while addressing vulnerabilities posed by quantum computing.

W3ID innovatively generates secure Digital Object Identifiers (DOIs) tailored for the decentralized Web 3.0 ecosystem. Unlike traditional DOI systems from the Web 2.0 era, W3ID leverages advanced cryptographic hashing and timestamping techniques to produce unique and tamper-resistant identifiers for digital assets. This enables decentralized data management, robust authentication, and secure peer-to-peer transactions.

The W3ID generation process involves:

1. Converting digital objects into a standardized format.
2. Timestamping for uniqueness.
3. Applying a SHA-256 cryptographic hash.

Additionally, W3ID employs a dual-key system for secure authentication, enhancing both public and private verification mechanisms. Our results demonstrate the effectiveness of the W3ID model in maintaining the integrity and traceability of digital assets, even within decentralized networks.

This approach positions the W3ID model as a revolutionary solution for digital identity and authentication. By providing a standardized framework for private data handling, it aligns with Web 3.0 principles of decentralization while offering scalability and robustness. The W3ID model is uniquely equipped to support the emerging universal global digital twin world by creating a secure, universal ID system.

To further enhance encryption strength and authentication integrity in the quantum computing era, W3ID incorporates an advanced security mechanism. By requiring quadruple application of SHA-256, with three consecutive matches for validation, the system expands the number of possibilities to 256^4, which is approximately 4.3 billion times the current SHA-256 capacity. This dramatic increase in computational complexity ensures that even advanced quantum computing systems would face significant challenges in executing brute-force attacks. W3ID redefines digital identity standards for Web 3.0 and the quantum computing era, setting a new benchmark for security, scalability, and decentralization in the global digital twin ecosystem.

## Introduction

Web 3.0 is the evolution of web utilization and interaction. It adopts a read-write-execute upgrade to the internet, offering enhanced data security and ownership[1]. Key features include the semantic web, artificial intelligence, 3D graphics, connectivity, ubiquity, distributed ledger technology, and smart contacts. Unlike previous versions of the web, Web 3.0 leverages Artificial Intelligence (AI) and Machine Learning (ML) to evaluate data, allowing for intelligent development and sharing of useful information based on user preferences[2]. Historically, users had no control over how their data was stored or managed, as corporations frequently monitored user data without knowledge or consent.

Decentralization is the defining parameter of Web 3.0. It allows internet users to transact business peer-to-peer, eliminating intermediates or controlling entities in the process. This leads to a greater focus on user privacy, transparency, and ownership. Web 3.0 operates through decentralized networks of numerous peer-to-peer nodes (servers), creating an ecosystem of data interoperability.

Decentralized applications and blockchain technology, such as decentralized finance (DeFi), revolutionized the exchange of digital assets. DeFi offers transparent and trustworthy protocols that operate without intermediaries. By leveraging decentralized networks and open-source software, DeFi transforms traditional financial products into autonomous services. As noted, it aims to “reduce transaction costs, generate distributed trust, and empower decentralized platforms” by decentralizing the power of centralized institutions and fostering innovation. As such, DeFi expands access to financial services, providing entrepreneurs and innovators with the tools to develop customized financial solutions. Similarly, non-fungible tokens (NFTs) represent a new system of “how digital assets are marketed and monetized”. These tokens are unique digital identifiers bound to virtual properties, enabling new forms of ownership and authenticity in the digital world. They have opened up new economic possibilities, allowing creators and collectors to engage in decentralized, peer-to-peer transactions that redefine the exchange of digital content. By utilizing decentralized networks, NFTs demonstrate how true value can be established by the masses rather than central authorities. This approach means the worth of digital assets is determined by market demand, mirroring how cryptocurrencies derive their value.

In the context of Web 3.0, the adoption of digital object identifiers (DOIs) proves useful for the identification and authentication of digital objects. The W3ID model generates unique digital IDs for any digital object based on personalized information. Specifically, the model produces both a public and private ID, the former accessible to the public and the latter only to the user. This ensures that users can secure, trace, and organize their data and other digital assets.

According to Bruwer (2014)[3], Web 3.0 “facilitate[s] a worldwide data warehouse where any format of data can be shared and understood by any device over any network.” Understanding the role of digital object identifiers in a Web 3.0 platform is crucial, as the model’s ability to integrate and retrieve diverse data formats across different networks can lead to more efficient and comprehensive data utilization. Nonetheless, the significance of the W3ID model raises concern: how do decentralization and peer-to-peer transactions improve user privacy and transparency? Compared to Web 2.0, what are the cost-effective strategies of Web 3.0? Furthermore, despite the key role of decentralized networks in financial transactions, how will the value of digital assets be maintained and stabilized in a system without central authority? This paper seeks to answer these questions and highlight the significance of DOIs in Web 3.0.

In Web 2.0, a widely used identifier system for digital content is the DOI. Currently, DOIs are strings of numbers and letters mainly used to uniquely identify digital content such as research papers and journal articles. These DOIs contain metadata of the object, most commonly the URL where the object is stored. Researchers can access these materials by clicking on the DOIs if they are hyperlinked or manually resolving them in the browser by typing https://doi.org/ before the DOI. For example, the DOI 10.1000/182 can be resolved by going to https://doi.org/10.1000/182. DOIs have been widely used for accurately storing the locations of the materials for a long time even if the URL or the material changes.

As the internet transitions to Web 3.0, data will move away from large databases managed by big technology companies. Instead, they will be uploaded into a network of devices that give and receive data as they request it. Furthermore, W3IDs will be much more accessible for individual users than DOIs which are currently limited to trusted journals and mediating organizations.

## Methods

### 1. ID generation

Unique IDs for digital objects are crucial in the expanding digital age. The process of generating W3IDs ensures distinctiveness regardless of the file format. The following methodology outlines the steps taken to generate a W3ID involving converting objects into a standardized format, adding a unique timestamp, and then applying a cryptographic hash function to ensure the security and uniqueness of the ID.

#### 1.1. Input

To generate a W3ID, users first submit the digital object they want an identifier of. This submission can be text, image, video, or in any format. The model takes the hexadecimal representation of the byte data of the object as all digital objects can be represented as such.

#### 1.2.Timestamp

To ensure that every identifier is unique, the model will also record the time at which the digital object was submitted in the time format YYYYMMDDHHMMSSffffff where:

- YYYY is the year
- MM is the month
- DD is the day
- HH is the hours in military time
- MM is the minutes
- SS is the seconds
- ffffff is the microseconds (μs)

An example timestamp in the format above is “20230503194715925404” which means the time the digital object was submitted to the W3ID model was the year 2023, May 3rd, 19:47 or 7:47 pm, 15 seconds and 925404 microseconds UTC.

**1.3.Concatenation**

The timestamp in the above format is then added to the beginning of the byte string of the submitted object.

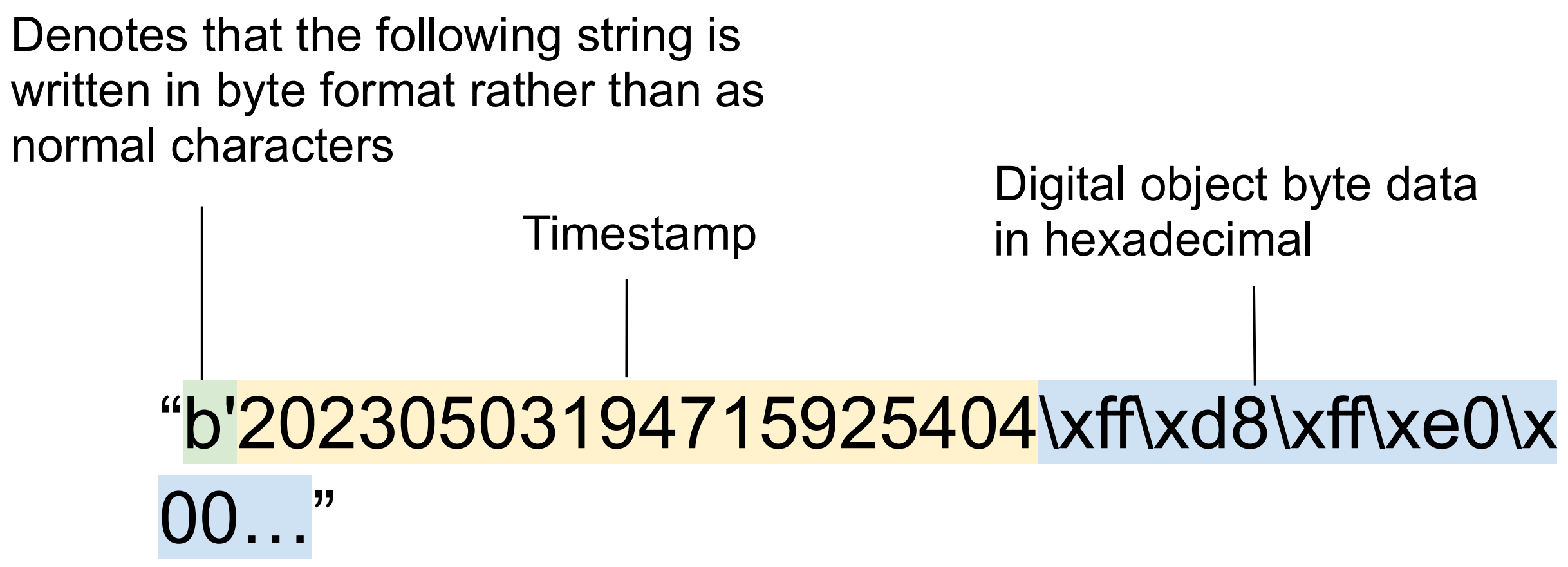

**Figure 1. An example of W3ID**

**1.4. Hashing**

This byte string is then sent through the SHA-256 hash function that outputs a scrambled version called a hash. The SHA-256 algorithm outputs a hash of 256 bits in length represented by 64 hexadecimal characters.

**1.5. Output**

“af9d89aa946b023f97e41623cb311bd5333685bc47904dd4089cd03ab8b2c49c” is given to the user as the object ID for their submission. This output can be further transformed into a QR code for easier access.

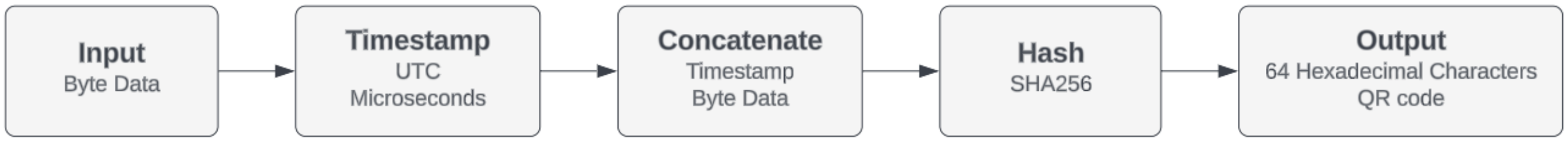

**Figure 2. The flowchart describes the process the ID is generated**. The input is taken as byte data, a timestamp accurate to microseconds, is attached to the byte data. Finally a SHA 256 hash is taken which outputs 64 hexadecimal characters and this can be easily stored through a QR code.

### 1.6. Example

Figure. 3 is (a) an example image and (b) an example pipeline of the W3ID model running (a). Figure 3 (a) is inputted into the W3ID model. The image is turned into byte data and is represented by the byte string b'\xff\xd8\xff\xe0\x00\x10JFIF\x00\x01\x01\x00\x00\x01\x00\x01\x00\x00\xff\xdb\x00C\x00\x02\x01\x01\x01\x01\x01\x02\x01\x01\x01\x02\x02\x02\x02\x02\x04\x03\x02\x02\x02\x02\x05\x04\x04\x03\x04\x06\x05\x06\x06\x06\x05\x06\x06\x06\x07\t\x08\x06\x07\t\x07\x06\x06\x08\x0b\x08\t\n\n\n\n\n\x06\x08\x0b\x0c\x0b\n\x0c\t\n\n\n\xff\xdb\x00C\x01\x02\x02\x02\x02\x02\x02\x05\x03\x03\x05\n\x07\x06\x07\n\n\n\n\n\n\n\n\n\n\n\n\n\n\n\n\n\n\n\n\n\n\n\n\n\n\n\n\n\n\n\n\n\n\n\n\n\n…’. The byte data was chosen due to the fact that any digital objects can be turned into byte data including web 3.0 digital objects.

The second step is timestamp. An example timestamp for the image above is 20230503194715925404 which means the time the digital object was submitted to the W3ID model was the year 2023, May 3rd, 19:47 or 7:47 pm, 15 seconds and 925404 microseconds. The timestamp ensures that each image submitted will have a unique DOI. In preparation of the third step, the timestamp is turned into byte data as well.

In the third step, the byte data of example digital object, is concatenated with the example timestamp in byte data. The following byte string looks like: ‘b'20230503194715925404\xff\xd8\xff\xe0\x00\x10JFIF\x00\x01\x01\x00\x00\x01\x00\x01\x00\x00\xff\xdb\x00C\x00\x02\x01\x01\x01\x01\x01\x02\x01\x01\x01\x02\x02\...’.

The fourth step is hashing. The byte string from above is inputted into the SHA256 algorithm which outputs ‘af9d89aa946b023f97e41623cb311bd5333685bc47904dd4089cd03ab8b2c49c’. Figure 4 is a QR code representation of the example image from Figure 3. The QR code represents the DOI generated by the W3ID model for the digital picture object in Figure 3, which is a digital twin representation of the scenery at the moment of the picture being taken with the time stamp of digital registration.

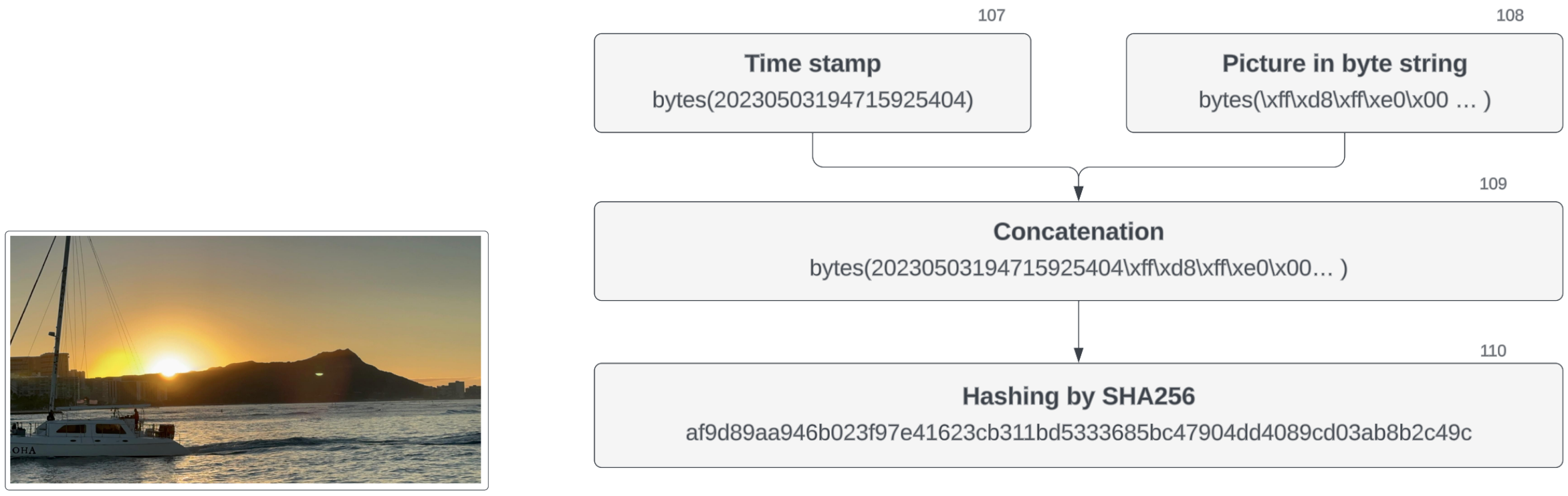

(a) (b)

**Figure 3. An example of W3ID on a digital picture with step-by-step implementation.** (a) Target digital object, and (b) step-by-step W3ID generation

**Figure 4. QR code of the W3ID in Figure 3.** The final W3ID from Figure3 is represented by a QR code.

## 2. Authentication

Authentication is a vital part of the digital world as objects can be easily fabricated and copied. By utilizing the uniqueness of the W3ID hash, we can create a trustworthy authentication system involving the splitting of private and public keys which can allow for public verification and private authentication.

### 2.1. Splitting the Hash

To enhance security in authentication systems, the 64-character hexadecimal hash can be split into two parts:

1. The first 32 characters act as the public key. This can be used to identify the digital object publicly
2. The last 32 characters act as the private key. This can be used to verify ownership or authenticity of the digital object.

For example, given the hash:

"af9d89aa946b023f97e41623cb311bd5333685bc47904dd4089cd03ab8b2c49c"

Public Key: af9d89aa946b023f97e41623cb311bd5

Private Key: 333685bc47904dd4089cd03ab8b2c49c

### 2.2. Authentication Process

Public Verification: When a digital object is accessed or shared, the public key can be used to verify its identity and make sure it wasn't tampered with. The public key should be accessible to everyone.

Private Authentication: To access the content linked to a W3ID, individuals would provide the private key or the full hash depending on the implementation. Only the author should receive the private key, although it is then up to the individual to ensure that the information is not leaked.

## 3. Quadruple extension of W3ID

To meet the challenges of the powerful quantum computing, we require four sets of W3ID to match. The four consecutive W3IDs are generated iteratively as follows. Starting with the first W3ID on a digital object, the second W3ID is generated from an updated time stamp of the new submission for the second W3ID. Then, this procedure will repeat for the third and the fourth of the W3ID's. Eventually, we have four consecutive sets of W3ID's with a required condition of the causality. After decoding, we can double-check the sanity of the four sets of W3ID's by checking its time progression of time stamps. This virtually make impossible to hack by algorithm, only by brutal random guessing. This will expand the number of possibilities to 256^4, which is approximately 4.3 billion times the current SHA-256 capacity, which makes even the quantum computing may not break it with brutal force approach.

## Results

### I. ID generation

The methodology described above provides a streamlined process with several different aspects. First, the inclusion of the timestamp ensures a unique digital ID for every digital object, even if it's the same object. Furthermore, the use of SHA256 standardizes the IDs into a string of 64 hexadecimal characters which gives it a consistent and manageable format for various applications such as searching in databases.

Additionally, while the methodology is fast and easy to implement, the ID provides security since due to the SHA256 algorithm, the hash is not reversible. Also since SHA256 generates extremely different IDs even with small changes can reveal any attempts of tampering or potential hacking. Lastly, the use of the timestamp allows the traceability of the original creator adding another layer of accountability.

### II. Authentication Application

The methodology for authentication uses the various advantages provided by the W3ID generation, primarily the fact that it is unique, to enable various processes like public verification and private authentication. Splitting the W3ID into two 32-character hexadecimal strings can allow for:

1. Public Verification: Using the public key for simple identification of the digital object.
    a. The public key could be implemented so that any user on the data storage platform can see the public key of any object. When users give and receive information, they can look at the public key or use an automated check to verify that the file they received is the one they requested.
2. Private Authentication: Using the private key to verify the authenticity or ownership of the digital object.
    a. An example use case is being able to access confidential information such as an email. If the private key does not match, then the user does not have the power to view the confidential email.

By doing so, the benefits of the W3ID system are maximized, ensuring that digital objects can be securely identified and authenticated across a wide range of systems.

## Discussion

Although the W3ID is designed to be integrated into the Web 3.0 landscape, it may be very useful in traditional, centralized data storage systems as a simple way to identify and keep track of data. Small companies could implement the W3ID when naming and manipulating data on their databases to ensure greater security, while individuals could also use it to name and store data in a standardized format.

The W3ID will become more valuable in the future where data is no longer the property of large technology companies and will take on many different formats. Due to the fact that the W3ID is able to generate unique and secure identifiers for any digital object makes it a promising way for individuals to name and organize data as it becomes more decentralized.

One way the W3ID would be valuable in Web 3.0 is as a standardized identifier in new P2P networks. While projects such as the InterPlanetary File System(IPFS) already use their own identifiers, a standardized system for identifying files across multiple platforms would be better than users having to keep track of different identifier formats for each data storage platform they use.

## Limitations and Future Directions

The main limitation of the W3ID model is that it currently needs to be more developed for use in Web 3.0 systems. The W3ID model is no more than an unsecured, unregistered website that generates an ID for something you submit. Because it isn't hosted on a secure server, it could run into problems such as the byte string exceeding the memory limit of the currently hosted system when trying to decode large files.

The W3ID itself is also not actionable; users cannot use the W3ID to visit or look at their data. As data will be stored in blockchains, the W3ID would become much more useful if it contained metadata about where data is stored in the chain, much like how DOIs in Web 2.0 contain the URL of the object they refer to. As it currently stands, however, this function would most likely have to be implemented by data storage providers rather than by the W3ID itself.

A potential development to the W3ID model is to set up a database of generated W3IDs along with the data storage platform they are part of. This would allow users to manually "resolve" W3IDs the way they can resolve DOIs and find where and what data is being stored with the given hash and ensure that the W3ID is not being used for encrypting harmful or illegal content. However, this development may contradict the decentralization aspect of Web 3.0 and would have to be implemented with caution.

An important future development to the W3ID could be to develop and release an API. The API would allow developers to easily implement the W3ID into their system by allowing them to automatically generate and assign W3IDs to objects. This functionality could also assist in managing the database mentioned above by automatically adding the generated W3ID along with some identifying information about which platform is hosting the data.

A feature of the W3ID that is distinct from DOIs is that they are guaranteed to be unique even if the submitted byte data is the same. This feature is not inherently a limitation as unlike research papers and journal articles, the W3ID aims to identify all types of digital content that may or may not be uploaded multiple times. Not only that, in peer-to-peer data storage networks it would mostly only harm users that upload large amounts of content as their device will take on the responsibility of hosting the data rather than a centralized server. Data storage providers would have to check for duplicate data before generating a W3ID if they wish not to store duplicates.

## Conclusion

In conclusion, the W3ID holds a lot of promise in both the current and future iterations of the internet as a secure, standardized, unique, and versatile identifier model for file data. Individuals and organizations can further utilize it for better authentication and data retrieval, which will especially benefit new P2P data storage services in the upcoming Web 3.0 environment.